\documentclass[12pt,english,floats,aps,amsmath,amssymb,nofootinbib,prd,shownopacs,floatfix]{revtex4-1}
\usepackage{babel}
\usepackage{amsmath}
\usepackage{graphicx}
\usepackage{amssymb}
\usepackage{ulem}
\usepackage{soul}
\usepackage{float}
\usepackage{subfig}
\usepackage{hyperref}

\usepackage{xcolor}
\definecolor{cblue}{RGB}{16,78,139}
\definecolor{cred}{RGB}{139,37,0}
\definecolor{cgreen}{RGB}{0,139,0}

\usepackage[left=2cm, right=2cm, top=2cm, bottom=2cm]{geometry}

\begin{document}

\title{Gauge-invariant bounce from loop quantum gravity}

\author{Klaus Liegener}
\email{liegener1@lsu.edu}
\affiliation{Department of Physics and Astronomy, Louisiana State University, Baton Rouge, LA 70803, USA}

\author{Parampreet Singh}
	\email{psingh@lsu.edu} 
\affiliation{Department of Physics and Astronomy, Louisiana State University, Baton Rouge, LA 70803, USA}


\begin{abstract}
We present a gauge-invariant treatment of singularity resolution using loop quantum gravity techniques with respect to local SU(2) transformations. Our analysis reveals many novel features of quantum geometry which were till now hidden in models based on non-gauge-invariant discretizations. Quantum geometric effects resolve the big bang singularity replacing it with a non-singular bounce when spacetime curvature reaches Planckian value. The bounce is found to be generically asymmetric in the sense that 
pre-bounce and post-bounce branches are not mirrored to each other and effective constants, such as Newton's constant, are rescaled across the bounce. Furthermore, in the vicinity of the bounce, minimally coupled matter behaves as 
 non-minimally coupled. These ramifications of quantum geometry open a rich avenue for potential phenomenological signatures.

\end{abstract}

\maketitle

Despite being extremely successful in describing the evolution of our universe from very early moments till present times, Einstein's theory of general relativity (GR) breaks down near the big bang singularity. It is expected that a quantum theory of gravity would provide valuable insights on singularity resolution.  In the last decade, techniques of loop quantum gravity (LQG) have been applied to study various cosmological spacetimes in loop quantum cosmology (LQC) \cite{Boj08, AS11}. The main result is the existence of a non-singular bounce, symmetric in absence of potentials and anisotropies,  due to non-perturbative quantum gravity effects occurring at Planckian spacetime curvature \cite{APS06a,APS06c,slqc}. 
Models including anisotropies and inhomogeneities have been explored, using an effective spacetime description, which extremely well captures underlying quantum dynamics \cite{numlsu-2}, and phenomenological implications for inflation and cosmic microwave background (CMB) have been investigated \cite{aan,as-review}. 

While this success is extraordinary, so far little is known about relating LQG with LQC, and the derivation of the cosmological sector of LQG is an open issue \cite{engle}. Further, some field theoretical aspects of LQG are not yet fully incorporated in LQC. LQG employs the insight of Ashtekar that a connection and triad based Hamiltonian formulation of GR is equivalent to a SU(2) Yang-Mills gauge theory. At the quantum level, smeared variables, i.e. holonomies of connection and fluxes of triads, play a central role. While the holonomies are inherently gauge-covariant, LQC uses a gauge fixing for the fluxes. Going beyond such a gauge fixing is important to verify whether physical predictions like the bounce are unaffected by gauge transformations. Else, a gauge dependence of fluxes can cause ambiguities, e.g. blurring even the differences between physical and degenerate metrics.  
A treatment based on gauge-covariant fluxes will not only provide unambiguous answers to whether a singularity is resolved, but may also reveal novel features which have so far been hidden in non-gauge-invariant approaches. This {\it{Letter}} aims to fill this important gap. Our treatment is based on the proposal for gauge-covariant fluxes introduced by Thiemann \cite{Thiemann:2000bv}. Details of this construction in the cosmological setting discussed in this manuscript can be found in our companion papers \cite{SL19b,LS19c}.

{\bf{{Limitations of conventional fluxes:}}}
The Ashtekar-Barbero connection $A^I_a$ and its canonical conjugate momentum $E^a_I$ for homogeneous, isotropic, spatially flat cosmology on a manifold $\sigma\cong \mathbb{R}^3$ are usually expressed using a gauge-fixing such that
\begin{equation}\label{FLRW-kinetic}
 A_a^I(x)= c \, \delta^I_a,\hspace{10pt} E^a_I(x)=p \, \delta^a_I
\end{equation}
where $c,p\in \mathbb{R}$, and the triad orientation is chosen to be positive. 
Though such a gauge-fixing, used in standard LQC, has the advantage of simplifying calculations in cosmology, it is nonetheless vital to construct {\it gauge-invariant} observables which are independent of the choice in (\ref{FLRW-kinetic}). We emphasize that here gauge invariance is meant  with respect to {\it all} gauge transformations $g(x)\in {\rm SU}(2)$ on $\sigma$. This includes  those transformations that introduce a dependence of $x\in\sigma$ on the RHS of (\ref{FLRW-kinetic}), i.e. do not leave the homogeneous subspace invariant. Such dependence though is artificial and by construction will not be seen via any gauge-invariant observable.

But constructing gauge-invariant observables in gauge theories is a non-trivial problem fraught with many difficulties. In these theories, one often requires an ultraviolet cutoff or a regularization parameter $\epsilon>0$ to avoid divergences during quantization. In particular, all observables built from continuum variables are discretized with respect to $\epsilon$ and it is not guaranteed that an arbitrary discretization yields a gauge invariant observable. For this, in fact, a careful construction is required after discretization. 

This problem directly affects LQG and hence LQC which is based on using the former's techniques in a cosmological setting with a non-vanishing discretization parameter $\epsilon$. Recall that in early works in LQG one introduces a  
discretization of the ``electric field'' $E^a(x):=E^a_I(x)\tau_I\in \mathfrak{su}(2)$, (with  $\tau_I = - i \sigma_I/2$ and $\sigma_I$ being the Pauli matrices) the {\it conventional fluxes}
\begin{equation}\label{mainstreamFlux}
E_I(S_a):=\int_{S_a} (* E)_I ,
\end{equation}
where for any $x\in \sigma$ a face $S_a$ (whose normal points in coordinate direction $a$) of coordinate area $\epsilon^2$ can be chosen, such that $\lim_{\epsilon\to 0} \epsilon^{-2}E_I(S_a)=E^a_I(x)$. Therefore, for any ${\rm SU}(2)$ gauge-invariant observable $O=O({E_I^a})$ there exists a discretized version $O^\epsilon=O^{\epsilon}({E_I(S_a)})$ built of conventional fluxes such that it reduces to the gauge-invariant observable in the limit $\epsilon\to0$. However, it is straightforward that for finite $\epsilon$ the quantity $O^\epsilon$ will in general {\it not} be gauge-invariant!

As an illustration, consider two triads, one for the spatially-flat isotropic cosmological model where $E^a_I$ is given by eq. (\ref{FLRW-kinetic}), and a {\it degenerate} metric $\tilde E^I_a(x)=p\;\delta^I_1(\delta^1_a+\delta^3_a)+p\;\delta_a^2\delta^I_2$. The function $Q(x) = \mathrm{det}(E)(x)$ is invariant  under $SU(2)$ gauge transformations $E^a(x) \mapsto \;^g E^a(x) =  g(x) E^a g^{-1}(x)$.
It has a conventional discretization
\begin{equation}
Q^\epsilon(x)=\frac{1}{48}\sum_{e_a\cap e_b\cap e_c =x}{\rm sgn}(\det(\dot{e}_a,\dot{e}_b,\dot{e}_c)) \epsilon^{IJK} E_I(S_a)E_J(S_b)E_K(S_c),
\end{equation}
where point $x$ is viewed as a six-valent vertex with incident edges $e_i$ having a constant tangent vector pointing in coordinate direction $i$.  At coordinate distance $\epsilon/2$ from $x$, $e_i$ is intersected transversal by a face $S_i$ 
which is oriented in direction $i$ and has coordinate area $\epsilon^2$.
	Taken together, all $S_i$ form a cube around $x$. 
	For $\epsilon \to 0$ the cube shrinks to $x$ yielding the continuum expression, i.e. $\epsilon^{-6}Q^\epsilon(x) \to Q(x)$.
	For a finite $\epsilon$, for the cosmological metric we find $Q^\epsilon[E](x)=\epsilon^6p^3$, and the degenerate metric gives $Q^\epsilon[\tilde E](x)=0$. 
	However, these results are gauge-dependent for any finite $\epsilon$. To see this, 
	consider a discretization with $\epsilon_0>0$, and a gauge transformation $g$ which is identity everywhere except in two small regions $N_+$ and $N_-$ which contain 
	 faces $S_{+1}$ and $S_{-1}$ of the cube respectively. Neither of $N_{\pm}$ contain parts of any other face nor $x$.  ($N_\pm$ can be considered a small disc of thickness $\delta < \epsilon$ with central surface $S_\pm$). 
If one would consider other discretizations $\epsilon \ll \epsilon_0$, then for $\epsilon\to 0$ the function $Q^\epsilon(x)$ will only see the trivial form of $g$ around $x$,
	i.e. its values for $E,\tilde E$ are unchanged.
	In the regions $N_\pm$, where $g$ is non-trivial, we choose $g$ such that it rotates the densitized triad via $g|_{N_\pm} =\frac{1}{\sqrt{2}}\binom{1\;\;-1}{1\;\;\;\;1}$. Then, one gets $Q^{\epsilon_0}[E](x) \mapsto {}^{g} Q^{\epsilon_0}[E](x) = 0$ for the cosmological metric, and $Q^{\epsilon_0}[\tilde E](x) \mapsto {}^{g} Q^{\epsilon_0}[\tilde E](x) = \epsilon^6_0 p^3$ for the degenerate metric! Thus, 
	this gauge transformation yields a physically inconsistent result for  a discretisation built with respect to the finite parameter $\epsilon_0$.

Therefore, using conventional fluxes with a finite regularization $\epsilon$, the discretization of $Q(x)$ capturing the volume of a cell around any $x\in\sigma$ does not distinguish whether the underlying metric is physical or degenerate resulting in a singular solution.  This is a serious drawback directly affecting LQC in a fundamental way since the latter uses conventional fluxes and non-vanishing regularization parameter \cite{APS06a,APS06c}.  The above example implies that singularity resolution is not guaranteed in models based on conventional fluxes. \\

{\bf{Gauge-covariant fluxes:}}
In LQG, for $\epsilon\to 0$  the SU(2) gauge-covariance of conventional fluxes is restored \cite{Thi98}. But, such a strategy does not work for LQC which is based on non-vanishing $\epsilon$. Thus, a gauge-covariant version of fluxes becomes necessary. Instead of (\ref{mainstreamFlux}) one uses a discretization of the densitized triads that transforms feasibly under ${\rm SU}(2)$-gauge transformations. The first proposal for a {\it{gauge-covariant flux}} was introduced by Thiemann \cite{Thiemann:2000bv}:
\begin{equation}\label{ThiemannFlux}
 P_I(x,S)= P_I(x,S)(A,E):=-2\; \mathrm{tr}\left(\tau_I h(e_{x,S})\int_S d^2y\; h(l_y)(* E)(y)h^\dagger(l_y) h^\dagger(e_{x,S}) \right)
\end{equation}
where $e_{x,S}$ is a path connecting $x$ with an interior point $x_0$ of $S$ and the paths $l_y$ connect $x_0$ with each $y$ 
while lying completely in $S$.

The $P(x,S)=P_I(x,S)\tau_I$ are {\it gauge-covariant} in the sense that $P(x,S) \mapsto g(x)P(x,S)g^{-1}(x)$. Thus, 
\begin{equation}
Q^\epsilon(x)=\frac{1}{48}\sum_{e_a\cap e_b\cap e_c =x}{\rm sgn}((\det(\dot{e}_a,\dot{e}_b,\dot{e}_c))  \epsilon^{IJK} P_I(x,S_a)P_J(x,S_b)P_K(x,S_c)
\end{equation}
is gauge-invariant for all ${\rm SU}(2)$-transformations and simultaneously for finite $\epsilon$. Explicitly, for the example discussed above this implies that $Q^{\epsilon_0}[E](x)= \;^gQ^{\epsilon_0}[E](x)\neq Q^{\epsilon_0}[\tilde{E}]$ for all gauge transformations $g$. Incorporating these gauge-covariant fluxes in the Hamiltonian constraint results in non-trivial modifications to the dynamics in contrast to using conventional fluxes. Even for the simplest models in LQC, it brings drastic changes in Planck regime physics as is discussed below.

{\bf{Quantum cosmological model:}}

The gauge-covariant fluxes (\ref{ThiemannFlux}) for the spatially-flat isotropic and homogeneous cosmological model can be found using coherent states methods in full LQG: Consider a coherent state $\Psi\in \mathcal{H}_\Gamma$, the LQG-Hilbert space restricted to a cubic lattice $\Gamma$ with spacing $\epsilon$ as discretization of the fiducial cell used for quantization. In loop quantization, the Hilbert space over every edge $e\in \Gamma$ can be expressed as $\mathcal{H}_e=L_2({\rm SU}(2),d\mu_H)$ with $d\mu_H$ being the Haar measure. The coherent state $\Psi=\otimes_e \psi_e$ in $\mathcal{H}_\Gamma :=\otimes_e \mathcal{H}_e$ can now be peaked for each edge $e$ on the holonomies $h_e$, and for either version of the fluxes: $E(S_e)$ or $P(e):=P(e(0),S_e)$. If $e$ is oriented along direction $I$, one finds \cite{SL19b}
 \begin{equation}\label{CosmologyMainstreamFluxes}
  E(S_e)= p\; \epsilon^2 \tau_I,
 \end{equation}
 while for a suitable choice of paths $l_y$ (such that they are always oriented along the axes defined by the edges of the lattice $\Gamma$ and the face $S_e$ is punctured by $e$ in the middle with respect to the fiducial coordinate system) in eq. (\ref{ThiemannFlux}) we obtain\footnote{Choosing different types of paths can in principle change the exact form of the following modification. However, it is to be noted that irrespective of this choice one expects non-trivial  gauge-covariant flux modifications from eq.(\ref{CosmologyMainstreamFluxes}) as discussed below.}
 \begin{equation}\label{CosmologyThiemannFluxes}
  P(e)= p\; \epsilon^2\; {\rm sinc}^2(c\epsilon/2) \tau_I ~.
 \end{equation}

To understand the physical implications of using gauge-covariant fluxes, we 
compute the expectation value in $\Psi$ of the Hamiltonian constraint. The classical Hamiltonian constraint in terms of Ashtekar-Barbero variables is given as
\begin{equation}
C=C_E-(\gamma^2+1)C_L + C_M
\end{equation}
with
\begin{align}
C_E&=F^I_{ab}\epsilon^{IJK}\frac{E^a_JE^b_K}{\sqrt{|{\rm det}(E)|}},\\
C_L&=\epsilon_{IMN}K^M_aK^N_b\epsilon^{IJK}\frac{E^a_JE^b_K}{\sqrt{|\det(E)|}}
\end{align}
corresponding to Euclidean and Lorentzian parts respectively, and $C_M$ denotes the matter part. Here, $F$ denotes the Lie algebra valued curvature of the connection $A$,  $K$ is the extrinsic curvature and $\gamma$ is the Barbero-Immirzi parameter. 
Following the procedure in LQG \cite{Thi98}, a symmetry reduced quantization of cosmological spacetimes can be performed.
Currently, two quantization procedures have been used in the literature: 
The first uses the classical relation $K = \gamma^{-1} A$ valid only for spatially-flat spacetime, resulting in $C_E\propto C_L$. It combines Euclidean and Lorentzian parts before quantization resulting in standard LQC with a scalar constraint operator $\hat C$ leading to a second-order quantum difference equation parameterized by $\epsilon$ \cite{Boj08,AS11}.
The other, known as Thiemann regularization \cite{ma,DL17,DL17b,klaus}, treats Euclidean and Lorentzian terms independently using that in general $K=2/(\kappa \gamma^3)\{A,\{C_E,V\}\}$. This results in Thiemann regularization \cite{ma,DL17,DL17b,klaus} with non-trivial modifications of order $\epsilon$ from $\hat C$ and leading to a fourth-order quantum difference equation. 

An additional regularization ambiguity appears via the choice of regulator $\epsilon$. Two common choices in LQC are a constant $\epsilon$ \cite{APS06a}, and $\epsilon = \bar \mu := \sqrt{\Delta/p}$, with $\Delta=4\sqrt{3}\pi\gamma G \hbar$. Other choices are possible \cite{lattice}, but in standard LQC only the $\bar \mu$ regularization is found in some sense to be a unique choice with respect to diffeomorphism invariance \cite{CS08,engle}, and yielding physically consistent ultra-violet and infra-red limits \cite{CS08}. The situation is the same on using gauge-covariant fluxes \cite{LS19c}. In the following we only consider  $\bar \mu$ regularization which is incorporated after modifications due to gauge covariant fluxes $P(e)$ are included in the Hamiltonian constraint.

Let us now discuss some details of the quantum Hamiltonian constraint for the  spatially-flat isotropic and homogeneous cosmological model with a minimally coupled massless scalar field $\phi$. This model  has been quantized rigorously in LQC where  the big bang singularity is replaced by a symmetric bounce of the universe at a universal value of energy density \cite{APS06a,APS06c,slqc}, with post-bounce and pre-bounce branches at large volumes matched to two classical solutions with the {\it{same}} physical constants. The strategy to obtain the quantum Hamitonian constraint with gauge-covariant flux modifications follows similar tratement as in standard LQC. One starts with the LQC Hilbert space $\mathcal{H}:=L_2(\bar{\mathbb{R}},d\mu_{\rm Bohr})$ on which the volume acts by multiplication $\hat{V}|v\rangle=v|v\rangle$, but in the present case encounters non-trivialities associated with ${\rm sinc}(c\bar \mu/2)$ which is not an almost periodic function. It turns out that the relevant solutions matching with the classical theory at late times can only be found for $c\bar \mu\in[-\pi,\pi]$, and one can restrict the $\mathrm{sinc}$ function to a compact interval where it allows a Fourier expansion \cite{SL19b}:
\begin{align}
{\rm sinc}(b)^2=T_{\infty}(b),\;\;\; T_N(b)=\frac{a_0}{2}+\sum_{n=1}^Na_n\cos(\frac{nb}{2})
\end{align}
with Fourier coefficients $a_n\in\mathbb{R}$. It is now possible to quantize this expression via $\cos(x\,b)\to (\mathcal N ^x +\mathcal N^{-x})/2$ where we introduced the shift operator $\mathcal{N}^x|v\rangle=|v+x\rangle$.
Constructed in this way, $\hat T_{\infty}$ is indeed a well-defined operator on $\mathcal H$ (see \cite{SL19b} for details). It is now straightforward to construct the quantum evolution operator. For the standard LQC regularization with a minimally coupled scalar field, the evolution operator using gauge-covariant fluxes reads,
\begin{align}
\Theta^{}:=-2\hbar^{-2}\hat{T}_\infty \hat{V}^{1/2}\hat{C}^{}\hat{V}^{1/2}\hat{T}_\infty
\end{align}
with
\begin{align}
\hat{C}^{}&:=\frac{-3\hbar}{32\gamma\sqrt{\Delta}}\left(f_{v+2}\mathcal{N}^4-g_v1_{\mathcal{H}}+f_{v-2}\mathcal{N}^{-4}\right)\nonumber
\end{align}
and $g_v=f_{v+2}+f_{v-2}$ and $f_v=-|v|(|v+1|-|v-1|)$. The analysis of quantum operator simplifies on truncating the Fourier series to a small number of terms. This truncation is very reliable and even at five terms the relative error in computation of triads is less than $1\%$ in the quantum regime \cite{SL19b}. The truncation error quickly decreases on inclusion of more terms. The resulting quantum evolution operator is more non-local than in standard LQC, resulting in a higher order quantum difference equation whose order is determined by the number of terms in the truncation.  \\

\begin{figure*}
	\begin{centering}
		\subfloat[]{\includegraphics[width=0.4\textwidth]{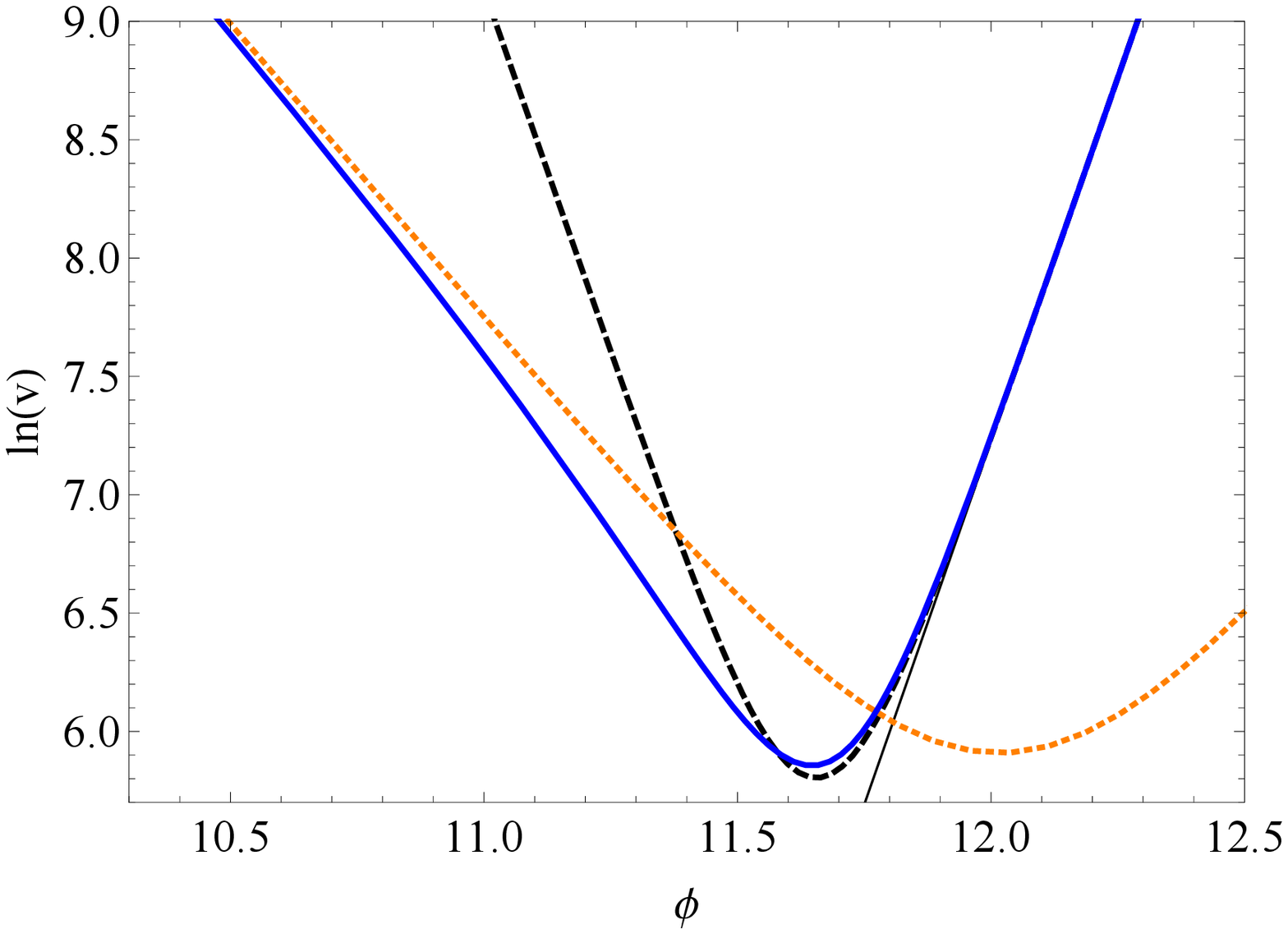}\label{figs1a}}
		\hspace{0,5cm}
		\subfloat[]{\includegraphics[width=0.395\textwidth]{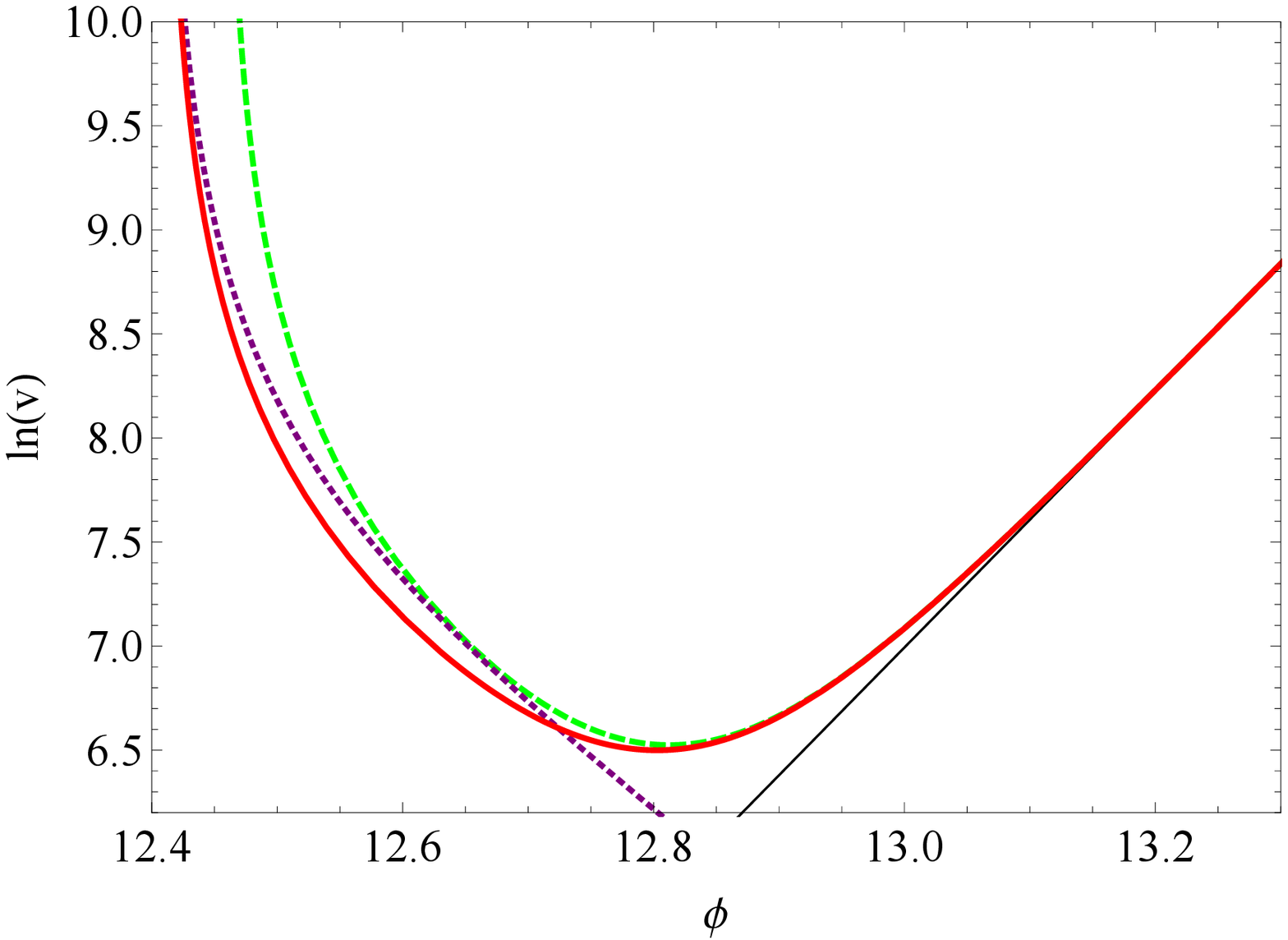}\label{figs1b}} 
	\end{centering}
	\caption{\small \raggedright Evolution of volume $v$ in clock-field time $\phi$ (choosing $\gamma=0.2375$ as is often done in LQC). On the left, we compare the volume in standard LQC (dashed black), with gauge-covariant volume (solid blue) and a rescaled LQC universe with $G',\pi_\phi'$ (dotted orange). In contrast to standard LQC, the bounce for gauge-covariant fluxes is asymmetric, even if ordinary volume is used. On the right, we show the Thiemann regularization adapted for LQC with standard fluxes (dashed, green), the new gauge-covariant fluxes (solid red) and a rescaled FLRW Universe with a positive cosmological constant $\bar{\Lambda}$ (dotted, purple).}
	\label{figs1}
\end{figure*}

{\bf{Effective spacetime description:}}
Extensive numerical evaluations confirm that the dynamics generated by LQC evolution operators can be well approximated by an effective Hamiltonian \cite{numlsu-2}. As is often assumed in standard LQC, we consider the effective spacetime description with a Fock quantized massless scalar field. In the following, we discuss the some main features of standard regularization followed by the ones of Thiemann regularization in LQC. For details, see \cite{SL19b,LS19c}.  \\
{\it{\ul{Standard LQC regularization:}}} The effective Hamiltonian constraint  (up to higher order $\hbar$ corrections) with lapse $N$ is \cite{klaus} 
\begin{align} \label{effectiveMainstreamHamiltonian}
C[N] = -\frac{3N \, p^{3/2}}{8\pi G\gamma^2\Delta}\sin^2(\bar{\mu} c){\rm sinc}(\dfrac{\bar{\mu} c}{2})
 ~+\frac{N\pi_\phi^2}{2 p^{3/2}}\;{\rm sinc}^{-3}(\dfrac{\bar{\mu} c}{2}) ~.
\end{align} 
Physical solutions, obtained from the vanishing of $C \approx 0$ and Hamilton's equations, turn out to be non-singular and exhibit a bounce of the gauge-covariant volume $v_{\mathrm{g.c}} :=  p \, {\rm sinc}(c\epsilon/2)$
in the Planck regime. One such solution is shown in Fig. 1 (left) where initial conditions are chosen at late times in a classical universe (identified by $\bar \mu c \rightarrow 0$). Using $\phi$ as a clock the evolution is approximated by classical GR (light-solid curve) till the spacetime curvature becomes $\sim 1 \%$ of its Planckian value, beyond which quantum gravitational effects become very significant.  The big bang singularity is resolved in backward evolution, a result which is independent of the choice of initial conditions. Unlike  standard LQC where the bounce is obtained using conventional fluxes, whose relation to the underlying geometry is blurred with respect to $SU(2)$ gauge transformations, {\it{the non-singular bounce found above is gauge-invariant}}. 
 
 The dynamics in the quantum gravitational regime and the nature of the bounce is different in non-trivial ways from standard LQC.
The energy density is bounded for physical solutions with maximum value $\rho_B \approx 0.028\gamma^{-2}$ (in Planck units). This is roughly 20\% larger than the value in standard LQC \cite{APS06c}. 
 But, this is least of the differences. Unlike LQC, the matter term in \eqref{effectiveMainstreamHamiltonian} has an explicit dependence on $\bar \mu c$ which is a function of spacetime curvature. Since, quantum geometry induces an explicit coupling of curvature with matter, {\it{ matter which is minimally coupled behaves as non-minimally coupled!}} The conservation law for energy density which held trivially in standard effective LQC, now holds if one carefully  takes the modifications to the dynamical evolution  due to gauge invariant fluxes  into account \cite{LS19c}.

Unlike the symmetric bounce in LQC, 
the gauge-invariant bounce turns out to be generally asymmetric. This is evident from Fig. 1 (left), where we have considered initial data in the expanding branch of a macroscopic universe. A comparison with the effective LQC trajectory shows an agreement far in the post-bounce regime, but a huge deviation in the pre-bounce regime. 
Analyzing the asymptotic behavior in the pre-bounce regime, we find that the trajectory indeed corresponds to a solution of classical Friedmann equations for vanishing spacetime curvature, albeit with a ${\it{rescaled}}$ Newton's constant $G \rightarrow G' = G[2/\pi]^4$. Moreover, from Hamilton's equation it transpires that the solutions  can be matched with systems with any rescaling of scalar field momentum by additionally rescaling the lapse appropriately \cite{SL19b}. However, like the lapse itself this rescaling does not contain physical information. (An analogous rescaling of the effective constants occurs for the expanding branch, if initial conditions are set at early times in the pre-bounce contracting universe). This rescaling is a ramification of gauge-covariant fluxes and is absent in standard LQC. The novel result is that 
{\it after passage through the quantum gravity region avoiding the singularity the asymptotic solution behaves like a classical one with changed constants of dynamics!} 

{\it{\ul{Thiemann regularization:}}}
Various qualitative features of bounce and Planck scale physics, such as non-minimal coupling and rescaling of effective constants, are robust  when using gauge-covariant fluxes in Thiemann regularization of the Hamiltonian constraint \cite{LS19c}: 
\begin{equation}\label{effectiveLorentzianHamiltonian}
C[N]=\frac{6\sqrt{p}^3}{\kappa\Delta}{\rm sinc}(c\bar{\mu}/2)(\sin^2(c\bar{\mu})+\frac{1+\gamma^2}{4\gamma^2}\sin^2(2c\bar{\mu}))+\frac{N\pi_\phi^2}{2p^{3/2}}{\rm sinc}^{-3}(\frac{\bar{\mu}c}{2})
\end{equation}
A key change occurs in the nature of the asymmetric bounce which occurs at bounce density 
\begin{equation}
\rho_B^{TR} = 0.023\gamma^{-2} {\rm Max}_{|b|<\pi}({\rm sinc}(b/2)^{-2}\sin(b)^2(1-(1+\gamma^{2})\sin(b)^2))
\end{equation}
and results in a fundamentally different dynamics in the pre-bounce regime (see Fig. 1 (right)). 
Quantum geometry effects result in an emergent cosmological constant of Planckian order in the pre-bounce regime, however with different values compared to the quantization based on triads \cite{DL17,Assanioussi:2018hee} where a rescaling of $G$ also occurred \cite{lsw}.
Unlike the case of \eqref{effectiveMainstreamHamiltonian}, the spacetime curvature in the pre-bounce regime is Planckian and in the large volume limit we find the following Friedmann equation
\begin{align}
	H^2|_{\rm{pre-bounce}} =  \frac{8\pi \bar{G}N^2}{3}\frac{\pi^2_\phi}{2p^3}+\frac{N^2\bar{\Lambda}}{3}+\mathcal{O}((\rho/\rho_B^{TR})^2)
\end{align}
and 
\begin{align}
\bar{G}&:= \nonumber G\; \frac{{\rm sinc^4}(\alpha/2)}{(\gamma^2+1)} \Bigg[1-5\gamma^2+5\gamma\left(\frac{1}{\alpha}-\frac{\cot(\alpha/2)}{2}\right)\Bigg]
\end{align}
where $H$ denotes Hubble parameter, $\bar{\Lambda} := 3\; \tfrac{{\rm sinc^2}(\alpha/2)}{(1+\gamma^2)^2 \Delta}$ and $\alpha:=\sin^{-1}(1/(\sqrt{1+\gamma^2}))$.

Independently of which of the two regularisations is used, the rescaling is fixed by the maximum value attained by  $c\bar{\mu}$ in the principal branch, and changes in other branches. 
Change in values of constants across singularities has been speculated earlier \cite{constant1}, and a model based on discrete quantum gravity discusses such a possibility \cite{constant2}. But, a key difference in our results from these studies is that the change in effective constants is {\it not random} but completely determined by non-singular dynamics. \\

{\bf{Phenomenological implications:}} Quantum gravitational effects in the pre-inflationary phase in standard LQC have been argued to leave potential signatures in CMB \cite{aan}. To capture them, the conservation of curvature perturbation at super-horizon scale, a result true for various classical gravitational theories, plays an important role. A priori this conservation is not guaranteed in presence of quantum geometric effects. With  gauge-covariant fluxes, the underlying assumptions leading to conservation of curvature perturbation are altered, and existence of such a quantity, which is often taken for granted in LQC, becomes non-trivial. It turns out that an analogous conserved quantity exists which  yields the standard curvature perturbation when quantum geometric effects become negligible. As a result, Planck scale physics can be probed via the CMB but there are non-trivial changes from the analysis of existing models in LQC. 

In all inflationary models studied in LQC using conventional triads quantum geometric effects enter via the change in the background gravitational dynamics with unchanged properties of minimally-coupled inflaton. Gauge-covariant fluxes affect these implications non-trivially. Not only the background dynamics is qualitatively different from standard LQC,  but the inflaton behaves as non-minimally coupled. Note that most minimally-coupled inflationary models, such as $\phi^2$ inflation, are tightly constrained by cosmic microwave background observations \cite{Planck}, since the predicted values of tensor-to-scalar ratio of perturbations are not in the favorable range. It has been argued that presence of even small non-minimal coupling can alleviate these constraints making the models such as $\phi^2$ inflation in agreement with observations. In particular, the modifications to cosmological perturbations equations resulting from non-minimal coupling mechanism results in a smaller tensor-to-scalar ratio \cite{ratio}. Since gauge-covariant flux modifications naturally result in a non-minimal coupling, they can potentially result in a lower  tensor-to-scalar ratio in comparison to the same model in standard LQC. Hence, in contrast to the standard LQC, gauge-covariant flux modifications are expected to be phenomenologically more favorable for models such as $\phi^2$ inflation. 
Further, given the highly asymmetric nature of the bounce, effects of the pre-bounce regime on perturbations, which have so far remained to be explored in LQC, are also expected to be non-trivial. Thus, the pre-inflationarly regime is drastically different from existing models in LQC and is expected to yield distinct signatures.

{\bf{Summary:}} 
Motivated by coherent state constructions, which are potentially useful for deriving the cosmological sector of LQG,
we have presented a first ever SU(2) gauge-invariant treatment of singularity resolution in LQC. Compared to standard LQC, our analysis uses gauge-covariant fluxes which provide unambiguous predictions and reveals various so far hidden features of underlying quantum geometry. On one hand, our analysis confirms the existence of a bounce, the main result of LQC, but shows that symmetric bounces in standard LQC are an artifact of the way fluxes are gauge fixed. Although invariance with respect to spatial diffeomorphisms is still an open question, SU(2) gauge-covariant fluxes lead generically to an asymmetric bounce both for standard as well as Thiemann regularization in LQC, with effective constants changing their values across the bounce. 
Our analysis is a non-trivial step of bringing LQC closer to LQG, but ultimately one has to address the problem of discretization ambiguities, which can be potentially solved using Hamiltonian renormalization methods \cite{LLT17}.
 However, it transpires already that the novel ramifications of quantum geometry, such as minimally-coupled matter behaving as non-minimally coupled, and non-trivial changes in dynamics due to gauge-covariant fluxes will change qualitatively the phenomenological 
 implications of Planck scale physics in the very early universe from all existing LQG inspired cosmological models.\\

\noindent\textbf{Acknowledgements:} This work is supported by NSF grant PHY-1454832.

%
%
\end{document}